\documentclass[fleqn,12pt,twoside]{article}
\usepackage{amsbsy}
\usepackage{amssymb}
\usepackage[headings]{espcrc1}

\readRCS
$Id: espcrc1.tex,v 1.2 2004/02/24 11:22:11 spepping Exp $
\usepackage{graphicx}
\usepackage{epsfig}
\usepackage[figuresright]{rotating}

\newcommand{\be}{\begin{equation}}
\newcommand{\ee}{\end{equation}}
\newcommand{ \bea }{\begin{eqnarray}}
\newcommand{ \eea }{\end{eqnarray}}
\newcommand{ \la }{\langle}
\newcommand{ \ra }{\rangle}

\newcommand{\AmS}{{\protect\the\textfont2
  A\kern-.1667em\lower.5ex\hbox{M}\kern-.125emS}}

\title{Forward-Backward Charge Fluctuations at RHIC Energies}

\author{
Stephane Haussler\address{
Frankfurt Institute for Advanced Studies (FIAS), J.W. Goethe Universit\"at, \\
Max von Laue Stra\ss{}e 1, 60438 Frankfurt am Main, Germany},
Mohamed Abdel-Aziz\address[MCSD]{Institut f\"ur Theoretische Physik, J.W. Goethe Universit\"at, \\
Max von Laue Stra\ss{}e 1, 60438 Frankfurt am Main, Germany
        }%
, Marcus Bleicher\addressmark.}

\begin{document}

\maketitle
\begin{abstract}
We use the ultra-relativistic quantum molecular dynamic UrQMD
version 2.2 to study forward backward fluctuations and compare our
results with the published data by the PHOBOS.
\end{abstract}
\vspace{0.6cm}
\section{Introduction}\label{sec:introduction}
One of the main goals of the relativistic heavy ion program is to
understand the nature of the hadron production mechanism (e.g.\
parton coalescence, string fragmentation or cluster decay). Recent
RHIC data suggested the formation of  a quark gluon plasma (QGP)
during the collision of two heavy gold nuclei at center of mass
energy $\sqrt{s_{NN}}=200$~GeV. Using correlations and
fluctuations to probe the nature of the created QCD matter has
been proposed by many authors, see for
example\cite{Haussler:2005ei}.
Recently, the PHOBOS experiment performed a similar analysis to
the UA5 experiment \cite{Alpgard:1983xp} for Au+Au reactions at
$\sqrt{s_{NN}}=200$~GeV \cite{Back:2006id}. In ref
\cite{Abdel-Aziz:2006fe}, a simple model was introduced to extract
the effective cluster multiplicity $K_{\rm eff}$ from PHOBOS data
and $K_{\rm eff}\sim2.7$ was found for peripheral collisions and
$K_{\rm eff}\sim 2.2$ for central collisions. The value of $K_{\rm
eff}$ in central collisions is close to the $pp$ value reported by
UA5 \cite{Alpgard:1983xp}. Note that all measured cluster
multiplicities $K_{\rm eff}$ are larger than the one that is
computed for a hadron resonance gas ($K_{\rm HG}=1.5$)
\cite{Stephanov:1999zu}, indicating that the measured charge
correlations can not be described by simple statistical models
based on hadronic degrees of freedom. Our goal in the current
study is to get a baseline estimate for forward-backward
fluctuations based on the microscopic hadronic transport model
UrQMD. For a complete review of the model see \cite{Bass:1998ca}.
In this paper we analyze $5\times10^5$ $pp$ and minimum bias Au+Au
events at $\sqrt{s_{NN}}=200$~GeV.
\section{Forward-Backward Fluctuations}\label{sec:model}
In this section, we introduce the variable $C$ that measures the
asymmetry between the forward and backward charges. We define two
symmetric rapidity regions at $\pm\eta$ with equal width
$\Delta\eta$. The number of charged particles in the forward
rapidity interval $\eta\pm\Delta\eta/2$ is $N_F$ while the
corresponding number in the backward hemisphere
$-\eta\pm\Delta\eta/2$ is given by $N_B$. We define the asymmetry
variable $C=(N_F-N_B)/\sqrt{N_F+N_B}$,
in each event. The variance of the charged particle multiplicity
in the forward hemisphere is given by $D_{FF}=\la N_F^2\ra-\la
N_F\ra^2$ and similarly for the backward hemisphere $D_{BB}=\la
N_B^2\ra-\la N_B\ra^2$. We also introduce the covariance of
charged particles in both hemispheres by $D_{FB}=\la N_FN_B\ra-\la
N_F\ra\la N_B\ra$, where $\la....\ra$ stands for the average over
all events. The PHOBOS measure of the dynamical fluctuations
$\sigma^2_C$ can be written as
\be \sigma^2_C\approx \frac{D_{FF}+D_{BB}-2D_{FB}}{\la
N_F+N_B\ra}.\ee

Recently STAR \cite{Tarnowsky:2006nh} reported a preliminary
results of the so called correlation strength parameter
$b=D_{FB}/D_{FF}$. The effective cluster multiplicity $K_{\rm eff}$ is propotional to
$\sigma^2_C$, such that if b=0, then the covariance $D_{FB}$ vanishes.
In this case we have $\sigma^2_C=K_{\rm eff}$.
We emphasize that $K_{\rm eff}$ should be understood as a product
of the true cluster multiplicity times a leakage factor $\xi$ that
takes into account the limited observation window $\Delta\eta$.
The event by event fluctuations of the asymmetry parameter
(variance) $\sigma^2_C$ in the absence of any correlations among
the produced particles will be $\sigma^2_C=1$.
%
%
\begin{figure}
 \vspace*{-0.0cm}
\centerline{
    \epsfig{figure=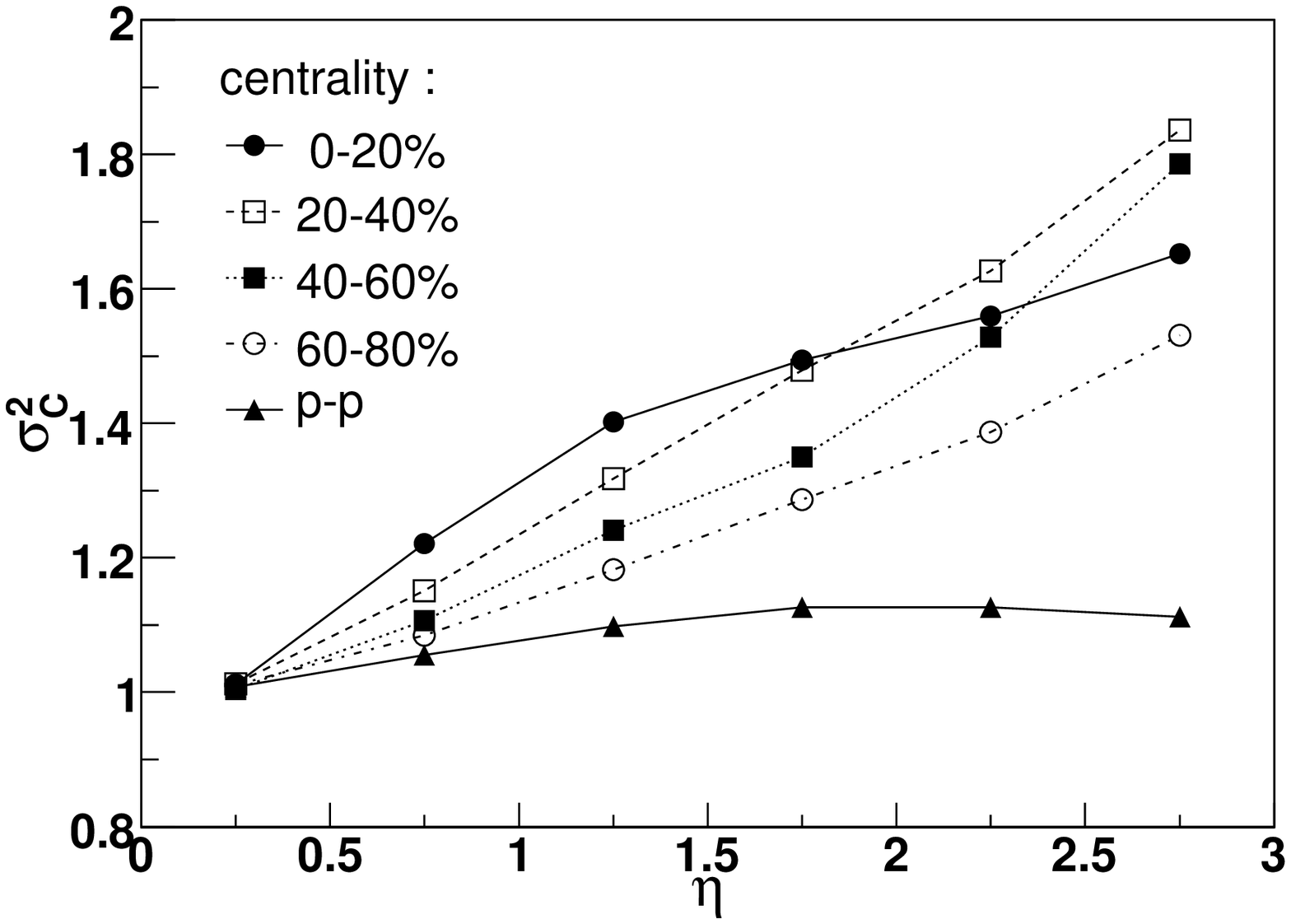,width=7.5cm}
    \hspace*{0.cm}
    \epsfig{figure=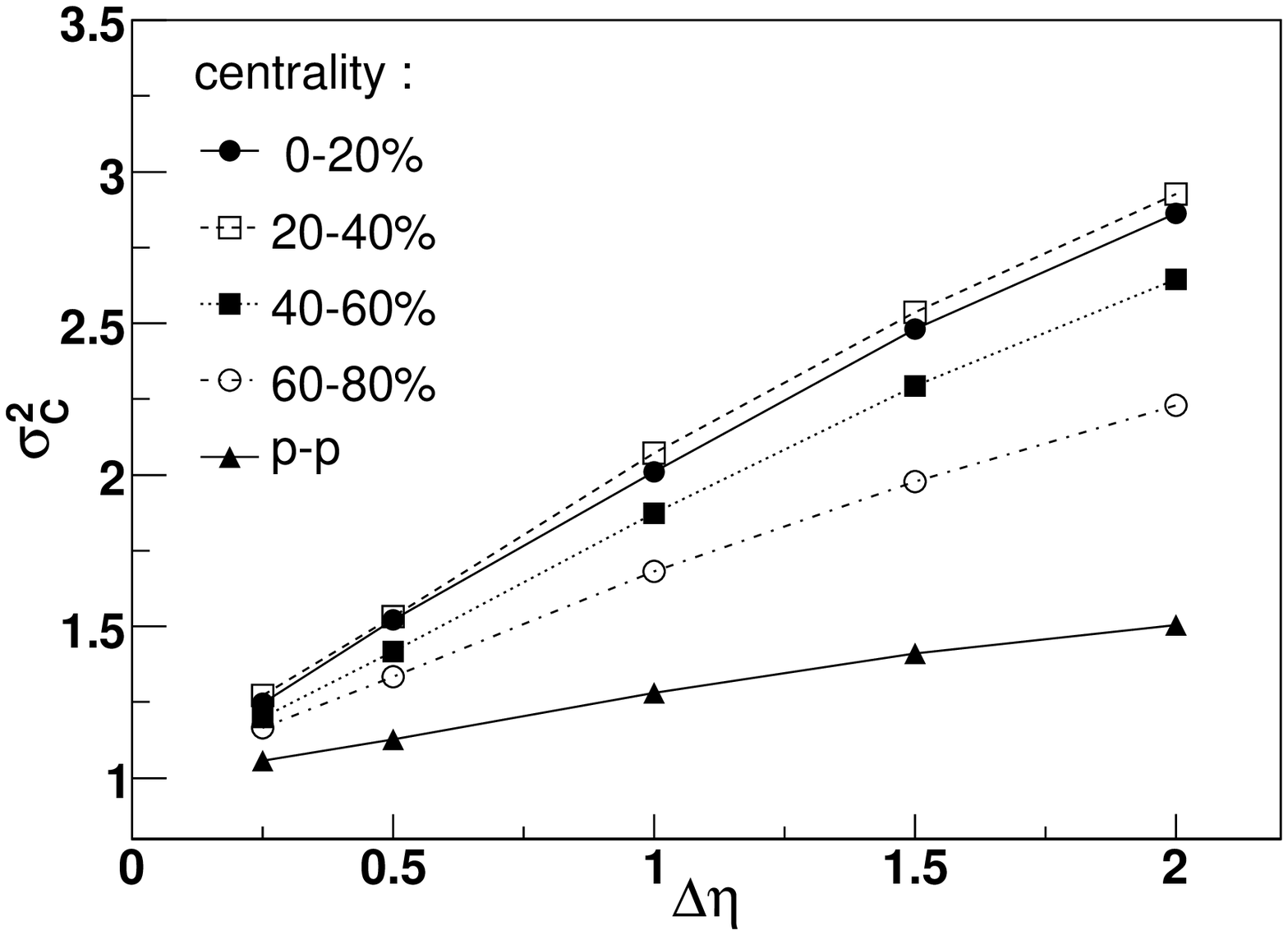,width=7.5cm}
} \vspace*{-0.95cm}
 \caption{UrQMD results of dynamic fluctuations as a function of $\eta$ with $\Delta\eta$=0.5 (left) and as a function of $\Delta\eta$ with $\eta$=2 (right).}
 \label{fig:1}
 \vspace*{-0.4cm}
\end{figure}
\section{Results}

\begin{figure}
 \vspace*{-0.0cm}
\centerline{
    \epsfig{figure=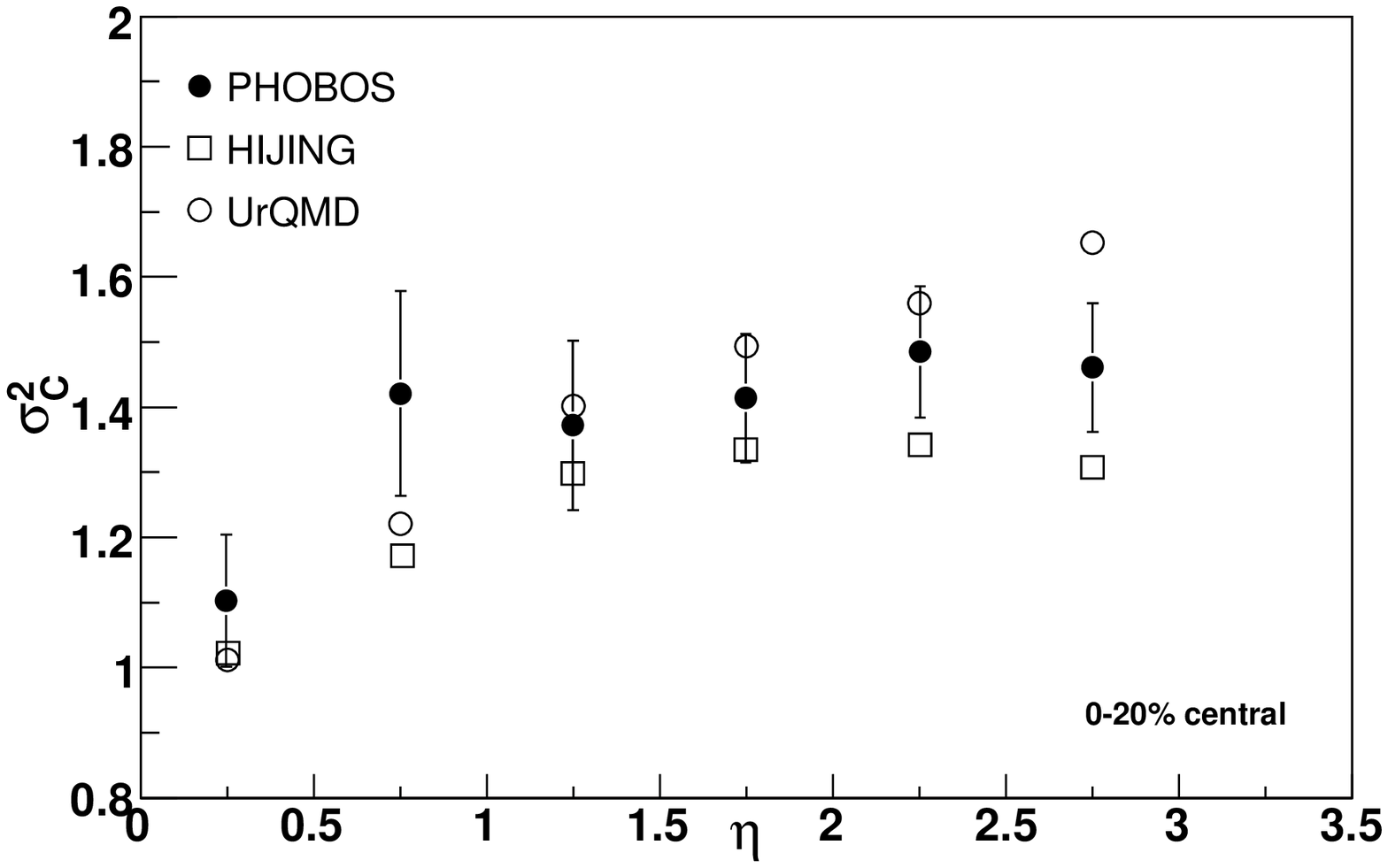,width=7.5cm}
    \hspace*{0.cm}
    \epsfig{figure=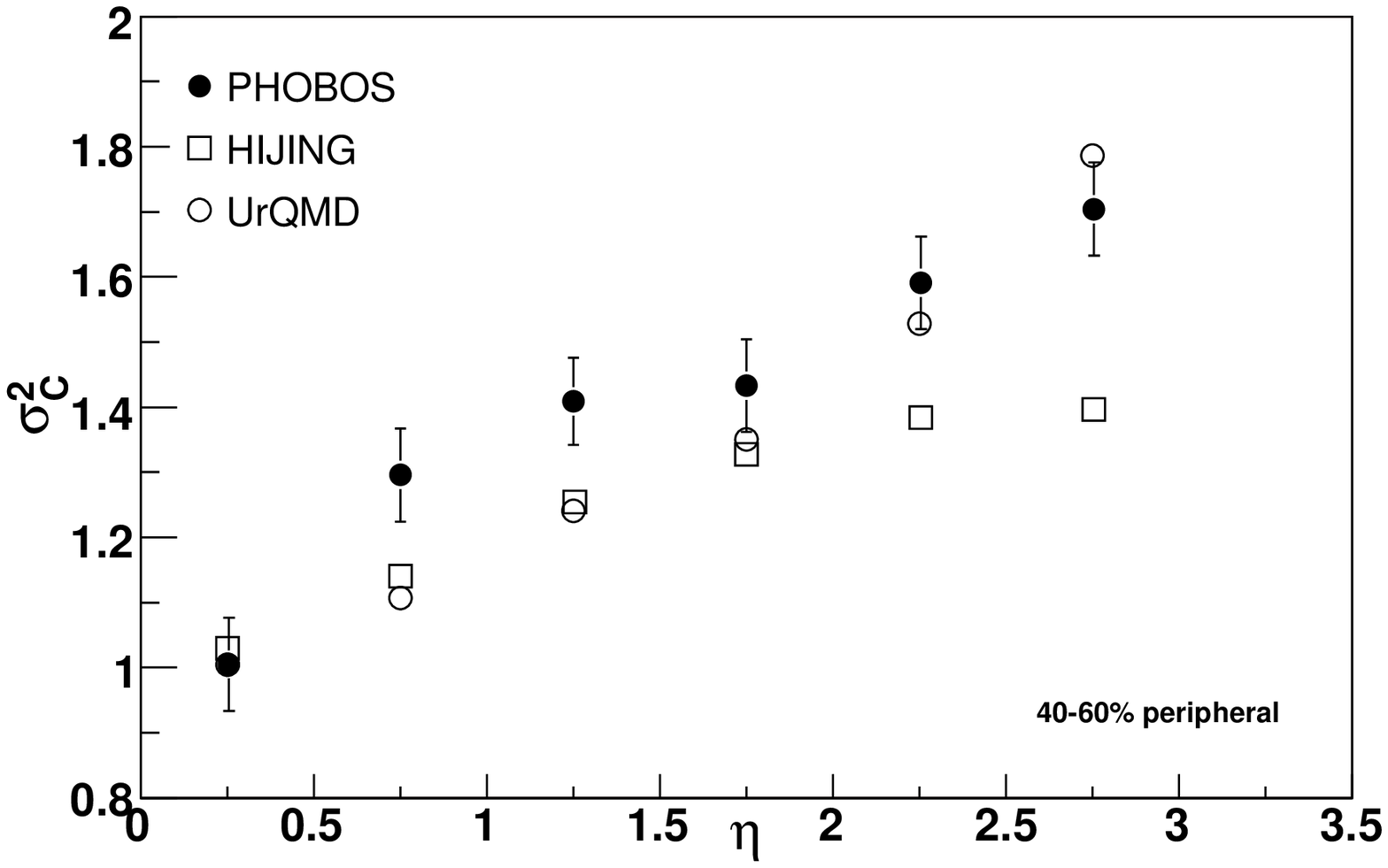,width=7.5cm}
} \vspace*{-0.95cm} \caption{$\sigma^2_C$ as a function of $\eta$ at $\sqrt{s_{NN}}=200$
 GeV, with $\Delta\eta$=0.5.
 (Left) $0-20\%$ central Au+Au, (right) $40-60\%$ peripheral Au+Au collisions.
 Black circles are PHOBOS \cite{Back:2006id} data, open squares HIJING and open circles are UrQMD results.}
 \label{fig:2}
  \vspace*{-0.4cm}
\end{figure}
\begin{figure}
 \vspace*{-0.0cm}
\centerline{
    \epsfig{figure=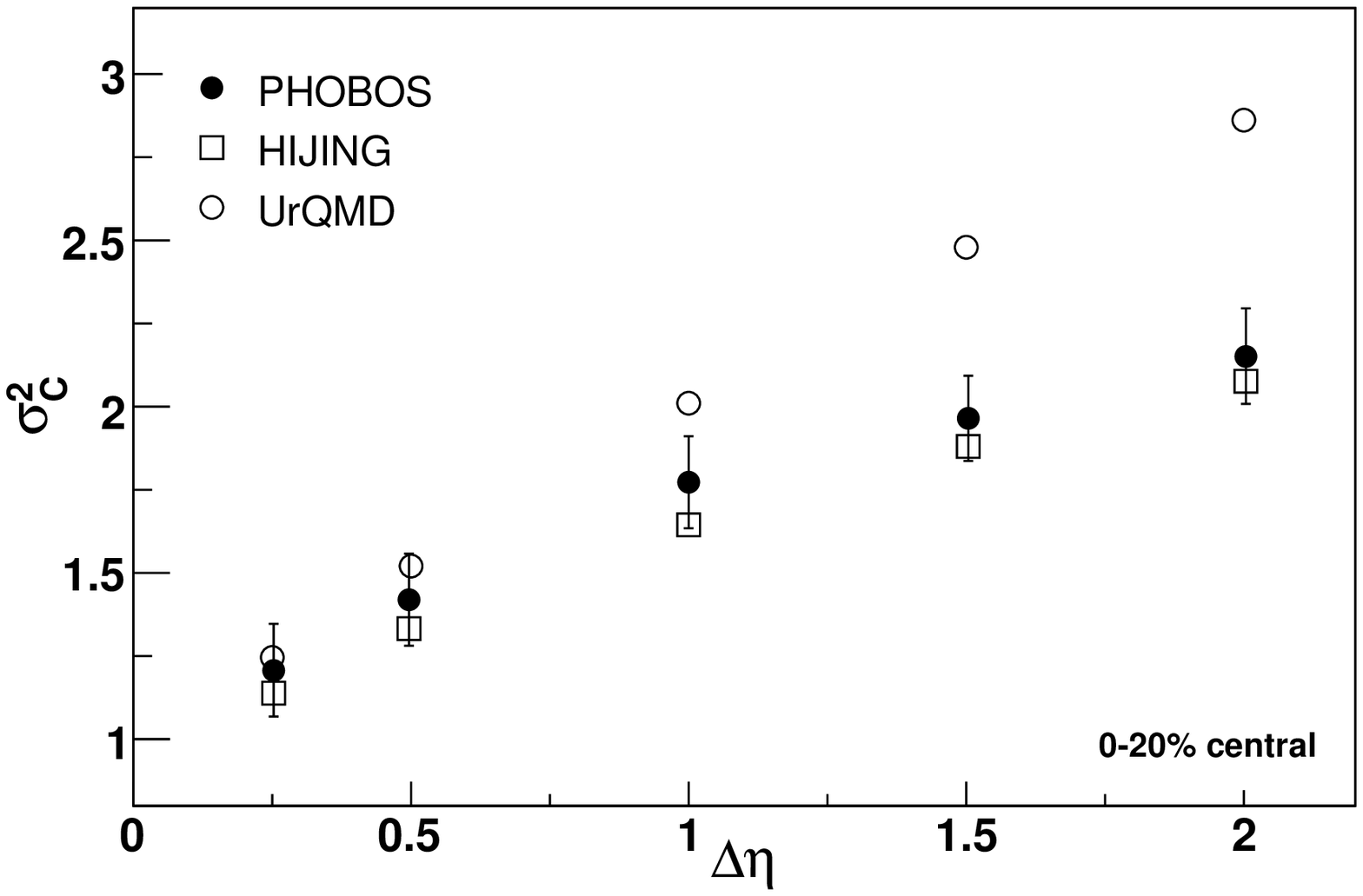,width=7.5cm}
    \hspace*{0.cm}
    \epsfig{figure=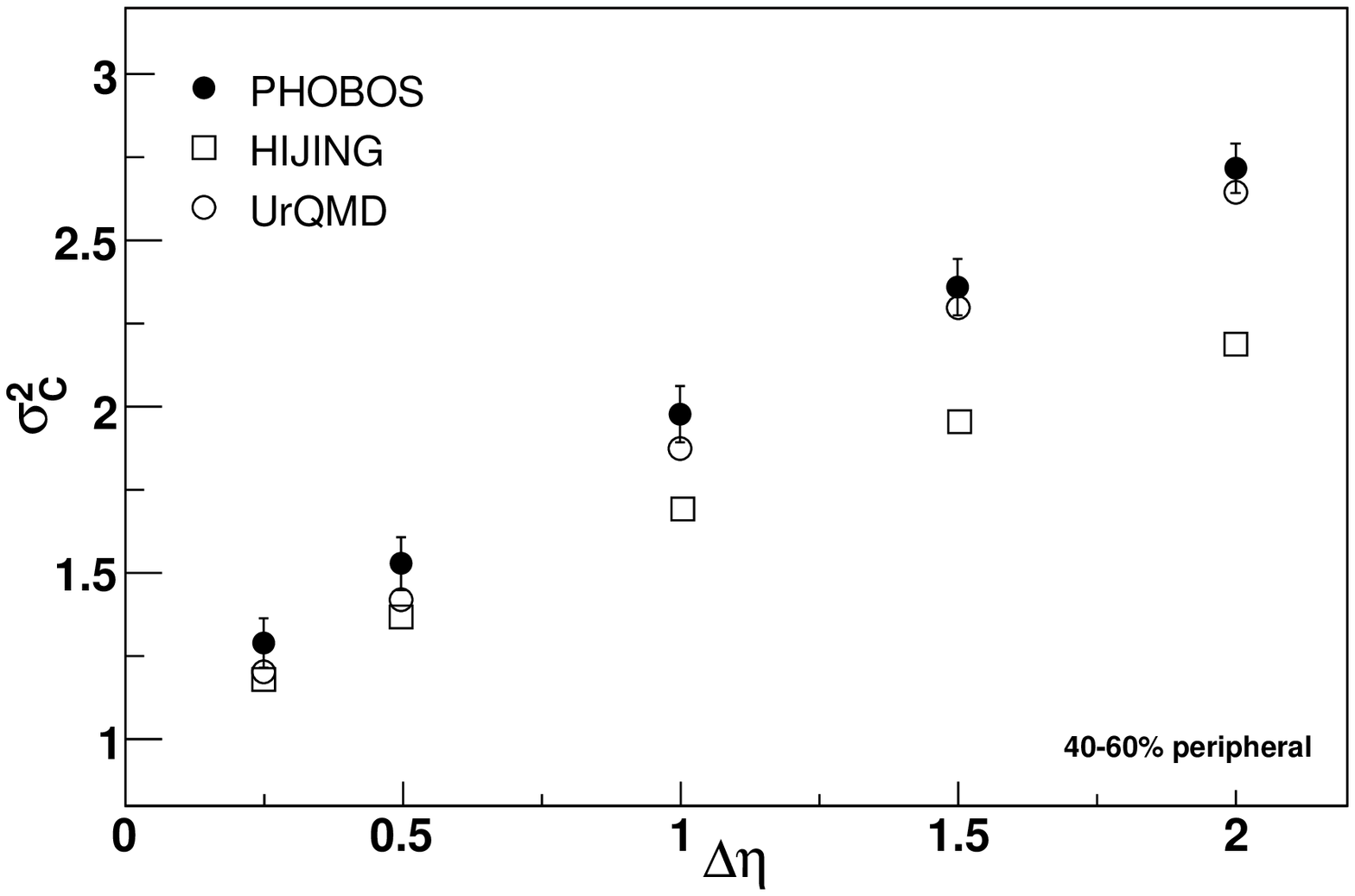,width=7.5cm}
} \vspace*{-0.95cm}
 \caption{$\sigma^2_C$ as a function of $\Delta\eta$ at $\sqrt{s_{NN}}=200$
 GeV, with $\eta$=2.
 (Left) $0-20\%$ central Au+Au, (right) $40-60\%$ peripheral Au+Au collisions.
 Black circles are PHOBOS data \cite{Back:2006id}, open squares HIJING and open circles are UrQMD
 results.}
 \label{fig:3}
  \vspace*{-0.4cm}
\end{figure}

In Fig.~\ref{fig:1}(left) we show $\sigma^2_C$ as a function of
$\eta$ at $\sqrt{s_{NN}}=200$~GeV computed from UrQMD. We find
that for $pp$ collisions $\sigma^2_C\approx1$ when $\eta=0.25$,
then $\sigma^2_C$ increases to 1.1 for $\eta=2.75$. In
Fig.~\ref{fig:1}(right) we plot $\sigma^2_C$ as a function of
$\Delta\eta$, we notice that $\sigma^2_C$ reaches approximately
1.6. This value is close to the resonance gas value $K_{\rm
eff}\sim 1.5$ mentioned in \cite{Stephanov:1999zu}.
In Fig. \ref{fig:1}, we plot UrQMD results for Au+Au at different
centralities. UrQMD shows that $\sigma^2_C$ as a function of
$\eta$ and $\Delta\eta$ has a clear centrality dependance, such
that $\sigma^2_C$ increases with centrality and then starts to
decrease with more centrality cuts. In \cite{Abdel-Aziz:2006fe} we
predicted that, for PHOBOS data, with more centrality cut,
$\sigma^2_C$ may be reduce to 1.9, and this reduction in
$\sigma^2_C\sim K_{\rm eff}$ may be regarded as an indication
for cluster melting at RHIC. In Fig.~\ref{fig:2} we show our UrQMD
with PHOBOS data \cite{Back:2006id} and their HIJING results for
$0-20\%$ central (left) and $40-60\%$ peripheral (right) Au+Au
collisions. In both figures, we keep $\Delta\eta=0.5$. From
Fig.~\ref{fig:2} we find that for both centralities $\sigma^2_C$
increases with increasing $\eta$. This behavior exists in both the
experimental data and HIJING. For both centralities we find that
$\sigma^2_C\approx 1$ when $\eta=0.25$. This is because the
competition between long and short range correlations almost
cancels. In $0-20\%$ central collisions Fig.~\ref{fig:2}, UrQMD
and HIJING roughly reproduce the experimental data within
1.5$\sigma$. In $40-60\%$ peripheral collisions
Fig.~\ref{fig:2}(right), UrQMD can roughly produce the PHOBOS data
while HIJING deviates by more than two $\sigma$ for the large
rapidity gaps.
In ref \cite{Abdel-Aziz:2006fe}, we show that by varying the
observation window $\Delta\eta$ one can see the whole cluster
structure. We keep the center of the observation window at
$\eta=\pm 2$ while we allow for the observation window to change.
In Fig.~\ref{fig:3} we show the PHOBOS data \cite{Back:2006id} in
addition to HIJING and our UrQMD analysis. We see that the
measured $\sigma^2_C$ increases up to 2.2 and 2.8 for central and
peripheral collisions respectively. In contrast to HIJING which
gives the same value for both centralities. UrQMD shows a
centrality dependance as shown in Fig \ref{fig:3}. For $0-20\%$
cental collisions, HIJING can reproduce the data. HIJING fails to
reproduce the the peripheral data. UrQMD over estimates
$\sigma^2_C$ in the central collisions while it succeeded to
reproduce the peripheral collisions. The failure of UrQMD to
reproduce the the central data indicated that the cluster
structure in UrQMD can survive, because the hadronic rscattering
is not strong enough to destroy such clusters.
\section{Summary and Conclusion}\label{sec:summary}
In this paper we computed forward-backward fluctuations and
compared UrQMD calculations results to the available experimental
data measured by PHOBOS. We started by studying proton-proton
collisions and we find that long range correlation persists over a
wide rapidity gap between the two rapidity hemispheres. The
variance of the asymmetry parameter $C$ was found to increase with
increasing $\Delta\eta$ such that $\sigma^2_C$ changes from
$\sigma^2_C(\eta=2,\Delta\eta=0.25)\approx 1$ to
$\sigma^2_C(\eta=2,\Delta\eta=3)\approx 1.6$, this can be due the
saturation of the leakage factor $\xi\rightarrow 1$.
For Au+Au collisions, we find that for both centrality bins
$0-20\%$ and $40-60\%$, $\sigma^2_C\approx 1$ for small
$\eta$. This can be seen as a cancellation between short and long
range fluctuations. By increasing $\eta$ we see that $\sigma^2_C$
also increases and approaches 1.6 and 1.8 for $0-20\%$ and
$40-60\%$ respectively. This increase can be attributed to the
decrease in the long range correlations. This will be true if the
particle production mechanism does not change with $\eta$. To see
the whole cluster structure, we fix the center of the observation
window at 2 and allow $\Delta\eta$ to increase. We find that UrQMD
can reproduce the peripheral data while it overestimates the
experimental results for central collisions. HIJING gives the
opposite behavior to UrQMD. HIJING produces the central data while
it fails to reproduce the peripheral data as shown by PHOBOS.
Also we find that UrQMD shows a smaller centrality dependance than
the data.
The discrepancies between HIJING, UrQMD and the data encourage
more theoretical study to be done in order to clarify the
correlation between produced particles in high energy nuclear
collisions.
The next step in this work is to measure $b$ and $\sigma^2_C$
consistently to extract the cluster multiplicity $K_{\rm eff}$ and
test for the survival or melting of such clusters.

\section*{Acknowledgements}This work is supported by BMBF and GSI.

\end{document}